\documentstyle[12pt]{article}
\newcommand{\nc}{\newcommand}
\nc{\postscript}[2] 
{\setlength{\epsfxsize}{#2\hsize}\centerline{\epsfbox{#1}}}
\nc{\bg}{B. Grzadkowski}
\nc{\non}{\nonumber}
\def\dps{\displaystyle}
\def\mib#1{\mbox{\boldmath $#1$}}
\def\sla#1{\mbox{$#1\!\!\scriptstyle{/}$}}

\def\bra#1{\langle #1 |} \def\ket#1{|#1\rangle}
\def\vev#1{\langle #1\rangle}

\nc{\barx}{\bar{x}}\nc{\pbarn}{\;\hbox {pb}}\nc{\fbarn}{\;\hbox {fb}}
\nc{\hc}{\hbox {h.c.}} \nc{\re}{\hbox {Re}} 
\nc{\mev}{\hbox {MeV}} \nc{\gev}{\;\hbox {GeV}}
\def\gesim{\lower0.5ex\hbox{$\:\buildrel >\over\sim\:$}} 
\def\lesim{\lower0.5ex\hbox{$\:\buildrel <\over\sim\:$}} 
\nc{\prd}[3]{{\it Phys.\ Rev.}\ {{\bf D{#1}} (#2), #3}}
\nc{\prl}[3]{{\it Phys.\ Rev.\ Lett.}\ {{\bf {#1}} (#2), #3}}
\nc{\plb}[3]{{\it Phys.\ Lett.}\ {{\bf B{#1}} (#2), #3}}
\nc{\npb}[3]{{\it Nucl.\ Phys.}\ {{\bf B{#1}} (#2), #3}}
\nc{\ptp}[3]{{\it Prog.\ Theor.\ Phys.}\ {{\bf {#1}} (#2), #3}}
\nc{\zfp}[3]{{\it Z.\ Phys.}\ {{\bf C{#1}} (#2), #3}}
\nc{\epj}[3]{{\it Eur.\ Phys.\ J.}\ {{\bf C{#1}} (#2), #3}}
\nc{\mpla}[3]{{\it Mod.\ Phys.\ Lett.}\ {{\bf A{#1}} (#2), #3}}
\nc{\rmp}[3]{{\it Rev.\ Mod.\ Phys.}\ {{\bf {#1}} (#2), #3}}
\nc{\ijmpa}[3]{{\it Int.\ J.\ of\ Mod.\ Phys.}\
               {{\bf A{#1}} (#2), #3}}
\nc{\ttbar}{t\bar{t}}         \nc{\bbbar}{b\bar{b}}
\nc{\tanb}{\tan \beta}        \nc{\twbdec}{t\to W^+ b}
\nc{\tbwbdec}{\bar{t}\to W^- \bar{b}}
\nc{\epem}{e^+e^-}            \nc{\eett}{\epem \to \ttbar}
\nc{\sigeett}{\sigma_{e\bar{e}\to\ttbar}}
\nc{\wpwm}{W^+W^-}            \nc{\tbar}{\bar{t}}
\nc{\bbar}{\bar{b}}           \nc{\wpp}{W^+}
\nc{\mt}{m_t}    \nc{\mts}{m_t^2}   \nc{\mw}{m_W}    \nc{\mws}{m_W^2}
\nc{\mz}{m_Z}    \nc{\mzs}{m_Z^2}
\nc{\ttbardec}{\ttbar \to W^+W^-\bbbar}
\nc{\wwbb}{W^+W^-\bbbar}      \nc{\sm}{SM}
\nc{\cw}{\cos\theta_W}        \nc{\sw}{\sin\theta_W}
\nc{\sws}{\sin^2\theta_W}     \nc{\sig}{\sigma_{tot}}
\nc{\lp}{{\ell}^+}              \nc{\lm}{{\ell}^-}
\nc{\epsl}{\epsilon_L}        \nc{\cp}{C\!P}
\nc{\splus}{s_+}       \nc{\smin}{s_-}        \nc{\eps}{\epsilon}
\nc{\psp}{Ps_+}        \nc{\psm}{Ps_-}        \nc{\lsp}{ls_+}
\nc{\lsm}{ls_-}        \nc{\sss}{s_+s_-}      \nc{\m}{m_t}
\nc{\mq}{m_t^2}        \nc{\mr}{\frac{1}{\m}} \nc{\av}{A_{\gamma}}
\nc{\bv}{B_{\gamma}}   \nc{\az}{A_Z}          \nc{\bz}{B_Z}
\nc{\avs}{A_{\gamma}^2}\nc{\azs}{A_Z^2}       \nc{\bzs}{B_Z^2}
\nc{\dav}{\delta \! A_{\gamma}}   \nc{\dbv}{\delta \! B_{\gamma}}
\nc{\dcv}{\delta C_{\gamma}}      \nc{\ddv}{\delta \! D_{\gamma}}
\nc{\daz}{\delta \! A_Z}          \nc{\dbz}{\delta \! B_Z}
\nc{\dcz}{\delta C_Z}             \nc{\ddz}{\delta \! D_Z}
\nc{\dev}{\delta \! E_{\gamma}}   \nc{\dez}{\delta \! E_Z}
\nc{\dfv}{\delta \! F_{\gamma}}   \nc{\dfz}{\delta \! F_Z}
\nc{\rdav}{{\rm Re}(\delta \! A_{\gamma}) \:}
\nc{\rdbv}{{\rm Re}(\delta \! B_{\gamma}) \:}
\nc{\rdcv}{{\rm Re}(\delta C_{\gamma}) \:}
\nc{\rddv}{{\rm Re}(\delta \! D_{\gamma}) \:}
\nc{\rdaz}{{\rm Re}(\delta \! A_Z) \:}
\nc{\rdbz}{{\rm Re}(\delta \! B_Z) \:}
\nc{\rdcz}{{\rm Re}(\delta C_Z) \:}
\nc{\rddz}{{\rm Re}(\delta \! D_Z) \:}
\nc{\idav}{{\rm Im}(\delta \! A_{\gamma}) \:}
\nc{\idbv}{{\rm Im}(\delta \! B_{\gamma}) \:}
\nc{\idcv}{{\rm Im}(\delta C_{\gamma}) \:}
\nc{\iddv}{{\rm Im}(\delta \! D_{\gamma}) \:}
\nc{\idaz}{{\rm Im}(\delta \! A_Z) \:}
\nc{\idbz}{{\rm Im}(\delta \! B_Z) \:}
\nc{\idcz}{{\rm Im}(\delta C_Z) \:}
\nc{\iddz}{{\rm Im}(\delta \! D_Z) \:}
\nc{\cz}{(1+v_e^2)d\:\!'^2}         \nc{\ci}{v_ed\:\!'}
\nc{\ccz}{v_ed\:\!'^2}              \nc{\cci}{d\:\!'}
\nc{\lspace}{\;\;\;\;\;\;\;\;\;\;}  \nc{\llspace}{\lspace \lspace}
\nc{\beq}{\begin{equation}}   \nc{\eeq}{\end{equation}}
\nc{\bea}{\begin{eqnarray}}   \nc{\eea}{\end{eqnarray}}
\nc{\baa}{\begin{array}}      \nc{\eaa}{\end{array}}
\nc{\bit}{\begin{itemize}}    \nc{\eit}{\end{itemize}}
\nc{\ben}{\begin{enumerate}}  \nc{\een}{\end{enumerate}}
\nc{\bce}{\begin{center}}     \nc{\ece}{\end{center}}
\nc{\ocal}{{\cal O}}
\begin{document}
\pagestyle{empty} \setlength{\footskip}{2.0cm}
\setlength{\oddsidemargin}{0.5cm} \setlength{\evensidemargin}{0.5cm}
\renewcommand{\thepage}{-- \arabic{page} --}
\def\mib#1{\mbox{\boldmath $#1$}}
\def\bra#1{\langle #1 |}      \def\ket#1{|#1\rangle}
\def\vev#1{\langle #1\rangle} \def\dps{\displaystyle}
\nc{\tb}{\stackrel{{\scriptscriptstyle (-)}}{t}}
\nc{\bb}{\stackrel{{\scriptscriptstyle (-)}}{b}}
\nc{\fb}{\stackrel{{\scriptscriptstyle (-)}}{f}}
\nc{\pp}{\gamma \gamma}
\nc{\pptt}{\pp \to \ttbar}
% -------------------------------------------------------------------
   \def\thebibliography#1{\centerline{REFERENCES}
     \list{[\arabic{enumi}]}{\settowidth\labelwidth{[#1]}\leftmargin
     \labelwidth\advance\leftmargin\labelsep\usecounter{enumi}}
     \def\newblock{\hskip .11em plus .33em minus -.07em}\sloppy
     \clubpenalty4000\widowpenalty4000\sfcode`\.=1000\relax}\let
     \endthebibliography=\endlist
   \def\sec#1{\addtocounter{section}{1}\section*{\hspace*{-0.72cm}
     \normalsize\bf\arabic{section}.$\;$#1}\vspace*{-0.3cm}}
% -------------------------------------------------------------------
\vspace*{-1cm}
\begin{flushright}
$\vcenter{
\hbox{IFT-41-01}
\hbox{TOKUSHIMA Report}
\hbox{(hep-ph/0112361)}
}$
\end{flushright}

\vskip 1cm
\begin{center}
{\large\bf Angular Distribution of Leptons in General}

\vskip 0.15cm
{\large\bf $\mib{t}\bar{\mib{t}}$ Production and Decay}
\end{center}

\vspace*{1cm}
\begin{center}
\renewcommand{\thefootnote}{\alph{footnote})}
{\sc Bohdan GRZADKOWSKI$^{\:1),\:}$}\footnote{E-mail address:
\tt bohdan.grzadkowski@fuw.edu.pl}\ and\ 
{\sc Zenr\=o HIOKI$^{\:2),\:}$}\footnote{E-mail address:
\tt hioki@ias.tokushima-u.ac.jp}
\end{center}

\vspace*{1cm}
\centerline{\sl $1)$ Institute of Theoretical Physics,\ Warsaw 
University}
\centerline{\sl Ho\.za 69, PL-00-681 Warsaw, POLAND}

\vskip 0.3cm
\centerline{\sl $2)$ Institute of Theoretical Physics,\ 
University of Tokushima}
\centerline{\sl Tokushima 770-8502, JAPAN}

\vspace*{2cm}
\centerline{ABSTRACT}

\vspace*{0.4cm}
\baselineskip=20pt plus 0.1pt minus 0.1pt
Angular distribution of the secondary lepton in top-quark production
followed by subsequent semi-leptonic decay is studied assuming
general top-quark couplings. It is shown that the distribution does
not depend on any possible anomalous $tbW$ couplings and is
determined only by the standard $V\!-\!A$ decay vertex for any
production mechanism if certain well-justified conditions are
satisfied.

\vspace*{0.4cm} \vfill

PACS:  14.60.-z, 14.65.Ha

Keywords:
anomalous top-quark interactions, lepton angular distribution \\

\newpage
%--------------------------------------------------------------------
\renewcommand{\thefootnote}{\sharp\arabic{footnote}}
%--------------------------------------------------------------------
\pagestyle{plain} \setcounter{footnote}{0}
\baselineskip=21.0pt plus 0.2pt minus 0.1pt
Ever since the top quark was discovered \cite{top}, a lot of data
have been accumulated. However, it still remains an open question
if the top-quark couplings obey the Standard Model (SM) scheme of the
electroweak forces or there exists a contribution from physics beyond
the SM. Although it is its heaviness that prevented us from
discovering this quark earlier, but once it is produced, the size of
the mass is a great advantage. Namely, the huge mass $m_t\simeq 174$
GeV leads to a decay width ${\mit\Gamma}_t$ much larger than $
{\mit\Lambda}_{\rm QCD}$. Therefore a top quark decays immediately
after being produced and the decay process is not influenced by
fragmentation effects \cite{Bigi}. This is why the decay products
could provide information on top-quark properties.

Next-generation $\epem$ linear colliders are expected to be a
top-quark factory, and therefore a lot of attention has been paid to
top-quark interactions in the process $\epem \to t \bar{t}$ (for a
review, see \cite{Atwood:2001tu} and the reference list there).
Although usually only anomalous $t\bar{t}\gamma/Z$ couplings have
been considered, however there is a priori no good reason to assume
that the decay part is properly described by the SM couplings. 
Therefore in a series of papers (see e.g.
\cite{Grzadkowski:1996kn,Grzadkowski:2000iq,Grzadkowski:2000nx}) we
have performed analyses of top-quark decay products assuming the most
general couplings both for the production and the decay.

In Ref.\cite{Grzadkowski:2000iq} we have noticed an amazing fact:
The angular distribution of the final leptons in $\epem \to t\bar{t}
\to {\ell}^{\pm}\cdots$ is not sensitive to modification of the
SM $V\!-\!A$ decay vertex. The same conclusion was also reached by
Rindani \cite{Rindani:2000jg} through an independent calculation
using the method of helicity amplitudes.\footnote{In
    Ref.\cite{Grzadkowski:2000iq} we have adopted the 
    Kawasaki-Shirafuji-Tsai formalism \cite{technique}, that will be
    also used here.}\
We usually suffer from too many parameters to be determined while
testing top-quark couplings in a general model-independent way.
Therefore, a distribution insensitive to a certain class of
non-standard form factors is obviously a big advantage as it
increases expected precision for the determination of other remaining
relevant couplings \cite{Grzadkowski:2000nx}.

In this short note we investigate if this interesting phenomenon
appears only in the process $\epem \to t\bar{t}\to {\ell}^{\pm}
\cdots$ or it could emerge within a wider class of processes. The
result is remarkable: It holds in quite a general context under some
natural and well-justified assumptions. In fact, we have observed
that the angular distribution of leptons from decays of polarized top
quark in its rest frame was free from the non-SM $tbW$ couplings
\cite{Grzadkowski:2000iq}. Since that was independent of a top-quark
production mechanism, it was already a strong indication that the
above decoupling would occur for any production process. Our goal
here is to provide the general proof for this hypothesis through
explicit calculations.

Let us consider a reaction like $1 + 2 \to \ttbar \to {\ell}^+ + X$
where the narrow-width approximation is applicable to the top
quark.\footnote{As the ratio of the top-quark width to its mass is of
    the order of ${\mit\Gamma}_t/m_t \simeq {\cal O}(10^{-2})$, the
    approximation is well justified.}\ 
We denote the momenta of the initial particles 1, 2 and the final
lepton as $\mib{k}_1$, $\mib{k}_2$ and $\mib{p}_{\ell}$,
respectively. For such processes, one can apply the
Kawasaki-Shirafuji-Tsai formula \cite{technique} in order to
determine the distribution of the final lepton:
\begin{equation}
\frac{d\sigma}{d\mib{p}_{\ell}} \equiv
\frac{d\sigma}{d\mib{p}_{\ell}}(1 + 2 \to \ttbar \to {\ell}^+ + X)
=2\int d{\mit\Omega}_t\frac{d\sigma}{d{\mit\Omega}_t}(s_t = n)
\frac1{{\mit\Gamma}_t}\frac{d{\mit\Gamma}}{d\mib{p}_{\ell}}.
\label{KST1}
\end{equation}
Here ${\mit\Gamma}_t$ is the top total width, 
$d{\mit\Gamma}/d\mib{p}_{\ell}$ is the spin-averaged top width
\[
\frac{d{\mit\Gamma}}{d\mib{p}_{\ell}} \equiv
\frac{d{\mit\Gamma}}{d\mib{p}_{\ell}}(t\to b{\ell}^+\nu)
\]
in the CM frame of $\ttbar$ pair,
and $d\sigma(s_t = n)/d{\mit\Omega}_t$ is the top-quark angular
distribution 
\[
\frac{d\sigma}{d{\mit\Omega}_t}(s_t = n) \equiv
\frac{d\sigma}{d{\mit\Omega}_t}(1 + 2 \to \ttbar \,;\:s_t=n)
\]
with its polarization vector $s_t$ being replaced with the so-called
``effective polarization vector" $n$
\begin{equation}
n_\mu = -\Bigl[\:g_{\mu\nu}-\frac{{p_t}_\mu{p_t}_\nu}{m_t^2}\:\Bigr]
{\sum\dps{\int} d{\mit\Phi}\:\bar{B}{\mit\Lambda}_+\gamma_5
\gamma^\nu B \over
\sum\dps{\int}d{\mit\Phi}\:\bar{B}{\mit\Lambda}_+ B},
\label{n-vec}
\end{equation}
where the spinor $B$ is defined such that the matrix element for
$t(s_t)\to {\ell}^+ +\cdots$ is expressed as $\bar{B}u_t(p_t,s_t)$,
${\mit\Lambda}_+\equiv \sla{p}_t +m_t$, $d{\mit\Phi}$ is the relevant
final-state phase-space element, and $\sum$ denotes the appropriate
spin summation.
 
Equation (\ref{KST1}) could be re-expressed in terms of the rescaled
energy and the direction of the lepton, $x$ and
${\mit\Omega}_{\ell}$:
\begin{equation}
\frac{d\sigma}{dx d{\mit\Omega}_{\ell}}
=2\int d{\mit\Omega}_t\frac{d\sigma}{d{\mit\Omega}_t}(s_t=n)
\frac1{{\mit\Gamma}_t}\frac{d{\mit\Gamma}}{dx d{\mit\Omega}_{\ell}},
\label{KST2}
\end{equation}
where $x$ is defined by the $\ttbar$ CM-frame lepton-energy
$E_{\ell}$ and $\beta\equiv\sqrt{1-4m_t^2/s}$ as
\[
x \equiv \frac{2E_{\ell}}{m_t}\sqrt{(1-\beta)/(1+\beta)}.
\]
It is natural to adopt $\mib{k}_1$ direction as the $z$ axis to
express $d\sigma/(dx d{\mit\Omega}_{\ell})$, while for the
width $d{\mit\Gamma}/(dx d{\mit\Omega}_{\ell})$, since it is
invariant under a three-dimensional orthogonal transformation,
we can use its form calculated in the frame where the
top-quark-momentum ($\mib{p}_t$) direction is chosen as the $z$
axis in the integrand on the right-hand side of eq.(\ref{KST2}).

The width in such a frame has been calculated in terms of $x$,
$\omega \equiv (p_t -p_{\ell})^2/m_t^2$ and the azimuthal
angle $\phi$ in \cite{Grzadkowski:1996kn}, assuming $m_{\ell}=
m_b=0$ and the most general decay couplings
\begin{eqnarray}
&&\!\!{\mit\Gamma}^{\mu}_{Wtb}=-{g\over\sqrt{2}}V_{tb}\:
\bar{u}(p_b)\biggl[\,\gamma^{\mu}(f_1^L P_L +f_1^R P_R)
-{{i\sigma^{\mu\nu}k_{\nu}}\over M_W}
(f_2^L P_L +f_2^R P_R)\,\biggr]u(p_t),\ \ \ \ \ \ \label{ffdef}\\
&&\!\!\bar{\mit\Gamma}^{\mu}_{Wtb}=-{g\over\sqrt{2}}V_{tb}^*\:
\bar{v}(p_{\bar{t}})
\biggl[\,\gamma^{\mu}(\bar{f}_1^L P_L +\bar{f}_1^R P_R)
-{{i\sigma^{\mu\nu}k_{\nu}}\over M_W}
(\bar{f}_2^L P_L +\bar{f}_2^R P_R)\,\biggr]v(p_{\bar{b}}),
\end{eqnarray}
where $P_{L/R}=(1\mp\gamma_5)/2$, $V_{tb}$ is the $(tb)$ element of
the Kobayashi-Maskawa matrix and $k$ is the momentum of
$W^{\pm}$ boson,\footnote{It is worth to mention that the form
    factors for top and anti-top quark satisfy
    \[
    f_1^{L,R}=\pm\bar{f}_1^{L,R},\lspace f_2^{L,R}=
    \pm\bar{f}_2^{R,L},
    \]
    where upper (lower) signs are those for $C\!P$-conserving
    (-violating) contributions \cite{cprelation}, assuming
    $C\!P$-conserving Kobayashi-Maskawa matrix. Therefore all the
    form factors contain both $C\!P$-conserving and $C\!P$-violating
    components. Since $W$ is on-shell, two extra form factors that
    are needed to describe the decay vertices do not contribute.}\ 
as
\begin{equation}
\frac{1}{{\mit\Gamma}_t}
\frac{d{\mit\Gamma}}{dxd\omega d\phi}
=\frac{1+\beta}{2\pi\beta}\;\frac{3 B_{\ell}}{W}
\omega\Bigl[\:1+2{\rm Re}(f_2^R)\sqrt{r}
\Bigl(\frac{1}{1-\omega}-\frac{3}{1+2r} \Bigr)\:\Bigr] \label{width}
\end{equation}
where $r \equiv (M_W/m_t)^2$, $B_{\ell} \equiv
{\mit\Gamma}/{\mit\Gamma}_t$, $W\equiv (1-r)^2(1+2r)$, $x$ and
$\omega$ are restricted as
\beq
0 \leq \omega \leq 1-r,\ \ \ \ 
1-x(1+\beta)/(1-\beta) \leq \omega \leq 1-x,
\label{omegalim}
\eeq
\beq
r(1-\beta)/(1+\beta) \leq x \leq 1.
\eeq
To find eq.(\ref{width}) we have 
assumed the standard $V\!-\!A$ coupling for $W\to{\ell}\nu_{\ell}$
and kept only SM contribution and the interference terms between the
SM and non-SM parts. Since we have neglected $b$-quark mass, only
$f_2^R$ interferes with the SM.

For eq.(\ref{ffdef}), the effective vector $n$ defined in
eq.(\ref{n-vec}) takes the following form \cite{Grzadkowski:2000iq}:
\begin{equation}
n^\mu =
\Bigl( g^{\mu\nu}-\frac{p_t^\mu p_t^\nu}{m_t^2} \Bigr)
\frac{m_t}{p_t p_{{\ell}^+}} (p_{{\ell}^+})_\nu. \label{n-vec2}
\end{equation}
It is worth to emphasize that this form is exactly the same as the
one given in the SM \cite{Arens:1992wh,Arens}. Namely $n^\mu$ did not
receive any non-standard corrections even though our calculation
assumed the most general top-quark decay vertex parameterization.

Changing one of the independent variables from $\omega$ to $
\theta$  (the angle between $\mib{p}_t$ and $\mib{p}_{\ell}$)
in the differential top-quark width (\ref{width}) through
\beq
\omega = 1-x{{1-\beta\cos\theta}\over{1-\beta}},
\label{omega}
\eeq
we have
\[
\frac{d{\mit\Gamma}}{dx d{\mit\Omega}_{\ell}}
=\frac{\beta x}{1-\beta} \frac{d{\mit\Gamma}}{dxd\omega d\phi}.
\]
Substituting this expression into eq.(\ref{KST2}), we are led to
\begin{eqnarray}
&&\frac{d\sigma}{dx d{\mit\Omega}_{\ell}}
=\frac{2 \beta x}{1-\beta}
\int d{\mit\Omega}_t\frac{d\sigma}{d{\mit\Omega}_t}(s_t=n)
\frac1{{\mit\Gamma}_t}
\frac{d{\mit\Gamma}_{\ell}}{dxd\omega d\phi} \non\\
&&\phantom{\frac{d\sigma}{dx d{\mit\Omega}_{\ell}}}
=\frac{3B_{\ell}}{\pi W}\frac{1+\beta}{1-\beta}\,x \non\\
&&\phantom{\frac{d\sigma}{dx d{\mit\Omega}_{\ell}}}
\ \ \ \ \ \times
\int d{\mit\Omega}_t\frac{d\sigma}{d{\mit\Omega}_t}(s_t =n)\,
\omega\Bigl[\:1+2{\rm Re}(f_2^R)\sqrt{r}
\Bigl(\frac{1}{1-\omega}-\frac{3}{1+2r} \Bigr)\:\Bigr].~~~~~
\label{KST3}
\end{eqnarray}
Once we have this formula, we may choose the lepton direction as the
$z$ axis for $d{\mit\Omega}_t$ integration. In this frame, the
top-quark polar angle $\theta_t$ is equivalent to $\theta$. So, in
the following, we will use eq.(\ref{omega}) with $\theta$ replaced by
$\theta_t$.

Let us derive constraints on $x$. Equation (\ref{omegalim}) implies:
\bit
\item 
$
0 \leq \omega \leq 1-r \; \Rightarrow \;
r(1-\beta)/(1-\beta\cos\theta_t) \leq x \leq
(1-\beta)/(1-\beta\cos\theta_t)
$
\item 
$
1-x(1+\beta)/(1-\beta) \leq \omega \leq 1-x \; \\
\phantom{-------} \Rightarrow \;
x \leq x(1-\beta\cos\theta_t)/(1-\beta) \leq x(1+\beta)/(1-\beta)
$
\eit
The latter constraint is trivially satisfied. So, when we perform $x$
integration first for a fixed $\theta_t$ in order to derive the
lepton angular distribution, its upper and lower bounds are
\begin{equation}
x_+ = (1-\beta)/(1-\beta\cos\theta_t),\ \ \ \
x_- = r(1-\beta)/(1-\beta\cos\theta_t).
\end{equation}

Here, whatever the top-quark production mechanism is,
$d\sigma(s_t=n)/d{\mit\Omega}_t$ depends on $p_{\ell}$ only through
$n$ vector and, however, {\it the effective vector $n$ has no
$x$-dependence in our case as it is directly seen in
eq.(\ref{n-vec2}) when the lepton mass is neglected.} Consequently
$d\sigma(s_t=n)/d{\mit\Omega}_t$ has no $x$ dependence at all and the
non-SM decay part of $d\sigma/d{\mit\Omega}_{\ell}$ is proportional
to
\[
\int^{x_+}_{x_-} dx\:
x\omega\Bigl(\frac1{1-\omega}-\frac3{1+2r}\Bigr).
\]
It is not hard to confirm that this integral vanishes. That is, the
non-standard-decay contribution disappears from the lepton angular
distribution for any top-quark production mechanism:
\begin{eqnarray}
&&\frac{d\sigma}{d{\mit\Omega}_{\ell}}
=\int dx \frac{x}{1-\beta}
\int d{\mit\Omega}_t\frac{d\sigma}{d{\mit\Omega}_t}(s_t=n)
\frac{1+\beta}{\pi}\;\frac{3 B_{\ell}}{W} \omega     \non\\
&&\phantom{\frac{d\sigma}{d{\mit\Omega}_{\ell}}}
=\frac{2 m_t^2 B_{\ell}}{\pi s}
\int d{\mit\Omega}_t \frac1{(1-\beta\cos\theta_t)^2}
\frac{d\sigma}{d{\mit\Omega}_t}(s_t=n),
\label{KST4}
\end{eqnarray}
where $d\sigma/d\mit\Omega_t(s_t=n)$ contains only information on
the production process. The last form of this equation is the
same as the one given by Arens and Sehgal within the SM
\cite{Arens:1992wh}.

Summarizing, we have shown that:\\
If the following conditions
\bit
\item the top-quark decay is described by the sequential processes
$\twbdec \to b {\ell}^+ \nu_l$,
\item narrow-width approximation is applied for $t$ and $W$,
\item only linear terms in non-standard form factors are kept,
\item $b$ quarks and final leptons are treated as massless,
\eit
are satisfied, then
\bit
\item linear corrections proportional to $f_2^R$ in the angular 
distribution of leptons
$$
\frac{d\sigma}{d\cos\theta_{\ell}}(1 + 2 \to \ttbar \to {\ell}^+ + X)
$$
vanish for any $\ttbar$ production process. So, only $V\!-\!A$
structure of the top-quark decay influences the leptonic angular
distribution.
\eit
There are a few comments in order here.
\bit
\item Non-standard effects are often parameterized in terms of $SU(2)
\times U(1)$ gauge symmetric, local and dim.6 effective operators
\cite{Buchmuller:1986jz,tree_oper}. Notice however that the above
theorem holds in a more general context than just the scenario of
effective operators: Since $\twbdec$ is a 2-body decay, all relevant
momentum products are fixed by the on-mass-shell conditions.
Therefore whatever the origin of $f_2^R$ and $\bar{f}_2^L$
is,\footnote{For instance, they could be loop-generated form factors.
    Notice that box-diagrams lead to 3-body decays and therefore such
    corrections are beyond our theorem, also emission of real photons
    or gluons is not included in our scheme. See \cite{jezqcd} for
    QCD corrections to the top-quark decays.} 
they are just constant numbers, and the proof goes through. Observed
deviation from the angular distribution, eq.(\ref{KST4}), could
indicate that $\twbdec$ is not the main decay channel of the top
quark.
\item An analogous conclusion applies also for the ${\ell}^-$ angular
distributions from $\tbar$ decays, i.e., disappearance of
$\bar{f}_2^L$.
\item As it was shown in \cite{Grzadkowski:2000iq}, the effective
polarization $n$-vector for the final $b$-quark distribution receives
an additional contribution from anomalous decay vertex and therefore
the angular distributions of $b$ quarks are sensitive to
modifications of the SM top-quark decay vertex in contrast to the
case of ${\ell}^\pm$.
\eit

In conclusion, we have proved that the lepton angular distribution in
the processes $1+2 \to \ttbar \to b{\ell}\nu_{\ell}X$ is independent
of any anomalous $tbW$ couplings regardless what is the production
mechanism. Therefore the distribution is sensitive only to
non-standard effects that enter the production process, and the
number of unknown top-quark couplings that parameterize the
distribution is reduced. We believe that for that reason the angular
distribution will be useful while testing top-quark couplings at
future colliders.

\vspace*{0.6cm}
% AAAAAAAAAAAAAAAAAAAAAAAAAAAAAAAAAAAAAAAAA
\centerline{ACKNOWLEDGMENTS}

\vspace*{0.3cm}

B.G. thanks Jos\'e Wudka for interesting discussions. This work is
supported in part by the State Committee for Scientific Research
under grant 5~P03B~121~20 (Poland) and the Grant-in-Aid for
Scientific Research No.13135219 from the Japan Society for the
Promotion of Science. 

\vspace*{0.6cm}
% RRRRRRRRRRRRRRRRRRRRRRRRRRRRRRRRRRRRRRRRRRRR

\end{document}